\documentstyle[twoside,fleqn,espcrc2,amsfonts]{article}

\newcommand{\AmS}{{\protect\the\textfont2
  A\kern-.1667em\lower.5ex\hbox{M}\kern-.125emS}}
\newcommand{\Up}{{U^\prime}}                
\newcommand{\dSB}{{\Delta S_B}}             
\newcommand{\ah}{{\hat a}}                  
\title{Stochastic Split Determinant Algorithms}

\author{Ivan Horv\'ath\address{Department of Physics, University of Virginia,
        Charlottesville, VA 22903, USA} 
        \thanks{Work supported in part by the DOE under
        grant No. DE-FG02-97ER41027.}}
       
\begin{document}

\begin{abstract}
I propose a large class of stochastic Markov processes associated 
with probability distributions analogous to that of lattice gauge
theory with dynamical fermions. The construction incorporates the
idea of approximate spectral split of the determinant through local 
loop action, and the idea of treating the infrared part of the split
through explicit diagonalizations. I suggest that {\it exact algorithms} 
of practical relevance might be based on the Markov processes so 
constructed.

\end{abstract}

\maketitle

One of the difficulties we face in lattice gauge theory is how to
efficiently simulate the gauge invariant distribution
\begin{equation}
   P(U) \; \propto\; e^{-S_B(U)} \det M(U)  \;\equiv\;
   P_1(U) P_2(U)
   \label{eq:6}
\end{equation}
with bounded bosonic action $S_B(U)$, and the fermionic kernel $M(U)$, 
such that $\det M(U) \ge 0$. 

The dynamic Monte Carlo approach, which is usually adopted, is 
based on identifying a suitable ergodic Markov process, defined 
by the Markov matrix $T(U,\Up)$, which has required distribution 
as a fixed point, i.e. $\int dU P(U) T(U,\Up) = P(\Up)$. While
finding examples of valid Markov processes is by no means difficult, 
finding a {\it suitable} one proved to be a formidable task. 

In the discussion that follows, I will construct Markov processes
built on rather different ingredients than those of established 
methods. They incorporate several interesting ideas that appeared 
in recent years, and the corresponding {\it exact algorithms}
have a chance of being reasonably practical, while perhaps 
being more accomodative of complicated actions of 
Ginsparg-Wilson type. The main influence in the construction I will 
present is the method used in the inexact Truncated Determinant 
Algorithm (TDA)~\cite{Dun99}.

Probability distribution that splits into two unnormalized products  
such as (\ref{eq:6}) can in principle be simulated using the following
statement. Let $T_1(U,\Up)$ be the Markov matrix satisfying detailed 
balance with respect to $P_1$, i.e.
$P_1(U)\, T_1(U,\Up)\, dU = P_1(\Up)\, T_1(\Up,U)\, d\Up$,
and let $P_2^{acc}(U,\Up) \;=\; \min [1, {P_2(\Up) \over P_2(U)}]$
be the Metropolis acceptance probability with respect to $P_2$.
Then the transition matrix 
\begin{equation}
   T(U,\Up) \;\equiv\; T_1(U,\Up) P_2^{acc}(U,\Up), \quad U\ne\Up
   \label{eq:11}
\end{equation}
satisfies detailed balance with respect to the distribution 
$P\propto P_1 P_2$. Consequently, $T$ represents a valid Markov process 
assuming $T_1$ is ergodic.

While there is a lot of freedom in choosing the process $T_1$, the 
algorithm based on $T$ very much exemplifies the difficulties associated
with simulating distribution (\ref{eq:6}). For one, the calculation of
acceptance probability requires the calculation of determinant ratio,
which is hard. Also, if $T_1$ represents the probabilities for the
sweep of suitable local changes, for example, then the determinant is
expected to fluctuate strongly, giving rise to severe problems with 
equilibration and, hence, useless algorithm. Nevertheless, this is
a natural starting point and the aim is to improve upon these bad things.

As a first point to notice, there is an additional significant freedom 
in the above procedure. In particular, it is possible to split the
distribution (\ref{eq:6}) in two parts in infinitely many ways by 
multiplying and dividing with arbitrary bounded positive function of the 
gauge fields, i.e.
\begin{equation}
    P_1(U) = e^{-S_B-\dSB} \quad
    P_2(U) = e^{\dSB} \det M
    \label{eq:16}
\end{equation}
Labeling the splits by $\dSB$, we thus have a family of Markov processes
$T[\dSB]$ constructed as in (\ref{eq:11}), each with the correct fixed
point.

The reason why this can be useful is implicitly contained in the well
established fact, that if the fermions are not very light, the effects 
of the determinant can be absorbed in the simple shift of the gauge 
coupling~\cite{deG94}. This suggests that one can follow the fluctuations
of the determinant in such a case by approximating it with a simple loop
action. Using such an action for the split $\dSB$ as described above can 
therefore presumably reduce the fluctuations of the determinant part 
substantially even if the fermions are not heavy. Analogous observations 
in more or less related contexts were made for example in 
Refs.~\cite{Irv97,Has99,deF98}.  

Supposing this works, we are still left with an awkward algorithm, because 
the evaluation of Metropolis acceptance probability is costly. One way to 
proceed is to consider the stochastic linear Kennedy-Kuti acceptance 
probability instead~\cite{Ken85}, which can be cheap for the price of 
introducing some extra noise.
Indeed, we can generalize the transition matrices (\ref{eq:11}) to
stochastic ones and use the following statement. Let $T_1(U,\Up)$
be the Markov matrix satisfying detailed balance with respect to $P_1$.
Let further ${\hat R}_a(U,\Up)$ be a stochastic estimator, depending
on the set of stochastic variables $a$, so that
$< {\hat R}_a(U,\Up) >_a \;=\; P_2(\Up)/ P_2(U)$, and let 
\begin{equation}
   {\hat P}^{acc}_{2,a} \;=\; 
   \cases{\lambda_+ \;+\; \lambda_- {\hat R}_a(U,\Up), &if $U>\Up$;\cr
          \lambda_- \;+\; \lambda_+ {\hat R}_a(U,\Up), &if $U<\Up$,\cr}
   \label{eq:21}
\end{equation}
where $\lambda_+,\lambda_-$ are constants and some ordering of the gauge
fields is assumed. If we define
\begin{equation}
  {\hat T}_a(U,\Up) \;\equiv\; T_1(U,\Up) {\hat P}^{acc}_{2,a}(U,\Up),
  \quad U\ne\Up
  \label{eq:26}
\end{equation} 
then $T(U,\Up) = < {\hat T}_a(U,\Up) >_a$ satisfies 
detailed balance with respect to $P \propto P_1 P_2$. Consequently, the
stochastic transition matrices (\ref{eq:26}) can serve to generate a valid
Markov sequence assuming that $T_1$ is ergodic and the individual
estimates ${\hat P}^{acc}_{2,a}$ can be interpreted as probabilities.   

With this framework in mind, we are now dealing with large class
of stochastic matrices ${\hat T}[\dSB,{\hat P}^{acc}_{2,a}]$ 
that are assigned to $P(U)$ and, in addition to the choice of split 
action, are also labeled by the choice of stochastic probability estimator.
The challenge now is to construct an estimator that is (a) reasonably    
cheap, (b) does not require too much noise so that the fluctuations
already reduced by the split will not come back, (c) introduces
negligible amount of probabilistic violations in the acceptance step.      

That this can be done in the useful novel way is suggested by the 
qualitative observation that the small eigenvalues of the lattice
Dirac operator contribute substantially to the long distance behaviour
of the effective fermionic action while large eigenvalues are more
relevant at short distances~\cite{Dun99}. This obviously makes very good
physical sense, but it should be said that to make more quantitative 
statements appears to be nontrivial. Nevertheless, it suggests that 
after the fluctuations of the full determinant are reduced by splitting
the distribution with some ultralocal loop action, what is left in the
determinant part is dominated by small eigenvalues. We  should thus 
construct an estimator ${\hat R}_a(U,\Up)$ entering (\ref{eq:21})
that assumes bulk contribution from small eigenvalues, thus being 
able to take advantage of the approximate spectral split.

Such an estimator can be constructed as follows. By definition, we 
need to estimate the quantity $R(U,\Up)=P_2(\Up)/P_2(U)$. 
If $\lambda_i$ denote the eigenvalues of $M(U)$, this can be written as
\begin{equation}
  {e^{\dSB(\Up)}\; \lambda_1^\prime \lambda_2^\prime \ldots \lambda_N^\prime
   \over
   e^{\dSB(U)}\; \lambda_1 \lambda_2 \ldots \lambda_N }
   \;\equiv\; x_1 x_2 \ldots x_N \;,
   \label{eq:31}
\end{equation}
where
$ x_i \equiv {\lambda_i^\prime \over \lambda_i}\; 
e^{[\,\dSB(\Up) - \dSB(U)\,]/N}$,
and the eigenvalues are assumed to be ordered by increasing magnitude.
The monomial can then be estimated by
\begin{eqnarray}
   \label{eq:36} 
   {\hat R}_a &=& x_1 \;+\; \ah_1 {x_1 \over \alpha_1}(x_2-1) \;+\; \\
              &+& \ah_1 \ah_2 {x_1 x_2 \over \alpha_1 \alpha_2} (x_3-1) 
                  \;+\; \ldots 
                  \nonumber \\
              &+& \ah_1 \ah_2 \ldots \ah_{N-1}
                  {x_1 x_2 \ldots x_{N-1} \over 
                   \alpha_1\alpha_2 \ldots \alpha_{N-1}} (x_N-1), 
                  \nonumber
\end{eqnarray} 
where $0 < \alpha_i \le 1$ and $\ah_i$ are independent random variables
distributed according to 
\begin{equation}
   {\hat a}_i \;=\; 
   \cases{1, &with probability $\alpha_i$;\cr
          0, &with probability $1-\alpha_i$.\cr}
   \label{eq:41}
\end{equation}
Obviously, one has $< {\hat R}_a >_a = x_1 x_2 \ldots x_N$ as desired, and 
it is useful to emphasize the following points: 

(A) It is trivial to generate $\ah_i$ by using the random number generator.
To get ${\hat R}_a$, one can first assign to it the value $x_1$ and generate
$\ah_1$. If $\ah_1$ is zero, this is the whole estimate because $\ah_1$
multiplies all the remaining terms. If $\ah_1$ is one, the second term
is added and $\ah_2$ is generated. Again, if $\ah_2$ is zero, the estimate
is completed and if it is one, the third term is added and $\ah_3$ generated 
etc. The calculation stops as soon as the first $\ah_i$ is zero, and only the
corresponding number of smallest eigenvalues is needed. The required number
of eigenvalues can be determined beforehand if desired, and the average 
number over many estimates can be tuned by changing $\{\alpha_i\}$.

(B) One can group the contributions of several eigenvalues into a
single variable $x_i$, and the number of eigenvalues so grouped does not
have to be the same for all $x_i$. For example, if the split action 
$\dSB$ is determined by fitting the ultraviolet part of the truncated
determinant~\cite{Dun99A}, then it is probably best to group into $x_1$
the smallest eigenvalues almost up to the truncation number, and 
group the remaining eigenvalues differently so that the stochastic
part of ${\hat R}_a$ does not fluctuate a lot. 

(C) Strictly speaking, $\lambda_i$ do not have to be the eigenvalues
of $M$. For example, in case of two flavours of Wilson fermions,
we can instead use $\lambda_i = \delta_i^* \delta_i$, where $\delta_i$
are the eigenvalues of one flavour operator. It is in fact desirable  
that the corresponding $x_i$ be real non-negative and so, if the individual 
eigenvalues are not, we assume that we can group them together
so that the resulting product is, or use additional properties such as
in the example above. This is possible in situations of practical interest.   

(D) The amount of noise introduced by the estimator (\ref{eq:36}) is
typically much less than with the traditional $e^{Tr \log M}$ estimators,
and can be tuned by changing the values $\{\alpha_i\}$.
This comes at the price of calculating the lowest eigenvalues.

For obvious reasons, it is natural to call the algorithms based
on ${\hat T}[\dSB,{\hat P}^{acc}_{2,a}]$ of (\ref{eq:26}) with
the estimator of type (\ref{eq:36}) the Stochastic Split Determinant
Algorithms (SSDA). They simultaneously use the complementary 
representations of effective fermionic action in terms of gauge loops
with increasing length (converging rapidly at small distances), and
in terms of Dirac eigenvalues with increasing magnitude (converging
rapidly at large distances). If the two representations overlap
sufficiently with small number of terms, then a simple split action
$\dSB$ can be found and $T_1$, ${\hat P}^{acc}_{2,a}$ easily adjusted, 
so that the number of probabilistic exceptions is negligible, 
and efficient {\it exact} SSDA results. 
The TDA work~\cite{Dun99,Dun99A} suggests that this might 
be the case for reasonably large lattices in $QCD_4$. The work on 
quantitative aspects of these statements is in progress and will be reported
elsewhere.

On sufficiently large lattices, algorithms like TDA or SSDA in their 
current form will eventually have inferior efficiency to that of HMC, which 
has more favourable volume scaling. However, the crucial advantage of TDA 
and SSDA is that they essentially treat fermions in the eigenspace of operator 
$M$. As such, they are in principle applicable to functions $f(M)$ as well. 
For example, simulations of two or arbitrary flavours of staggered fermions 
is straightforward here.

Finally, it should be emphasized that in the current context, the underlying 
process $T_1$ can always be chosen so that there will essentially be no 
exceptions at all. In bad cases this can rapidly increase the work per 
independent configuration, but exactness can always be achieved. 
However, the underlying philosophy appears to be sufficiently solid to
believe that the framework is large enough for some
practically relevant algorithms to be found here. 

I thank Hank Thacker for many pleasant discussions on the topics
presented here.

\end{document}